\documentstyle[amssymb,prb,aps]{revtex}
\begin{document}
\title{Orientational pinning and transverse voltage: 
Simulations and experiments in square Josephson junction arrays}
\author{V. I. Marconi, S. Candia, P. Balenzuela\cite{byline}
, H. Pastoriza, D. Dom\'{\i}nguez}
\address{
Centro At\'{o}mico Bariloche, 8400 San Carlos 
de Bariloche, Rio Negro, Argentina}
\author{P. Martinoli}
\address{
Institut de Physique, Universit\'e de Neuch\^atel, Rue A.-L. Breguet
1, CH-2000 Neuch\^atel, Switzerland}
\maketitle
\begin{abstract}
We study the dependence of the transport properties of square 
Josephson Junctions arrays with the direction of the applied dc current, 
both experimentally and 
numerically.  
We present computational simulations of current-voltage 
curves at finite temperatures for a single
vortex in the array ($Ha^2/\Phi_0=f=1/L^2$), and experimental measurements
in $100\times1000$ arrays under 
a low magnetic field corresponding to $f\approx0.02$. 
We find that the transverse voltage vanishes only in
the directions of maximum symmetry of the square lattice:
the [10] and [01] direction (parallel bias) and the
[11] direction (diagonal bias).
For orientations different than the symmetry directions, we find a
finite transverse voltage which depends strongly 
on the angle $\phi$ of the current. We find that vortex motion 
is pinned in the [10] direction ($\phi=0$), 
meaning that  the voltage response is
insensitive to small changes in the orientation of the current near
$\phi=0$.
We call this phenomenon orientational pinning. This leads
to a finite transverse critical current for a bias at $\phi=0$ and to
a transverse voltage for a bias at $\phi\not=0$.
On the other hand, for diagonal bias in the [11] direction
the behavior is highly unstable against small variations of $\phi$, 
leading to a rapid change from zero
transverse voltage to a large transverse voltage within a few degrees.
This last behavior is in good agreement with our measurements in arrays
with a quasi-diagonal current drive. 
\end{abstract}
\section{INTRODUCTION}

The interaction between the periodicity of vortex lattices (VL) and periodic
pinning potentials in superconductors has raised a great interest  both in
equilibrium systems \cite{martinoli,wire,jja,jjpin,dot,hole,yan,franz,nori0} and
in driven non-equilibrium systems.\cite{nori,nori2,md99,reich} Several
techniques have been used to artificially fabricate periodic pinning in
superconducting samples: thickness modulated  films,\cite
{martinoli} wire networks, \cite{wire} Josephson junction
arrays, \cite{jja,jjpin} magnetic dot arrays,\cite{dot} sub-micron hole
lattices \cite{hole}  and pinning induced by Bitter decoration. \cite{yan}

The ground states of these systems, which result from the competition
between the vortex-vortex  and the vortex-pinning
interactions, can be either commensurate or incommensurate vortex structures
depending on the vortex density.\cite{martinoli,nori0} These 
conmensurability effects in the ground state vortex configurations lead to
enhanced critical currents and resistance minima for the ``matching'' and
for the ``fractional'' (submatching)  vortex densities where the VL is
strongly pinned. At finite temperatures, it has been shown that there are
both a depinning and a melting transition, which can occur either
sequentially or simultaneously depending on the magnetic field. \cite{franz}

Very recently, the physics of driven vortices under periodic pinning has
been studied numerically both at zero temperature \cite{nori,nori2} 
and at finite temperatures.\cite{md99,reich} 
At $T=0$
there is a complex variety of dynamic phases. \cite{nori} At finite
$T$ there are two dynamic transitions when increasing
temperature at high drives: there is first a 
transverse depinning and second a melting transition of the moving vortex lattice.
\cite{md99}

Most of the effects of periodic pinning that have been studied are related
to conmensurability phenomena and the breaking of translational symmetry in these
systems. Less studied is the effect of the breaking of rotational symmetry
in periodic pinning potentials, in particular regarding transport
properties. One question of interest is how the motion of vortices changes
when the direction of the driving current is varied. If there is rotational
symmetry, the vortex motion and voltage response should be insensitive to
the choice of the direction of the current. However, it is clear that in a
periodic pinning potential the dynamics may depend on the direction of the
current. For example, in square Josephson junction arrays (JJA) it has been
found that the existence of fractional giant Shapiro steps (FGGS) depends on
the orientation of the current bias. When the JJA is driven in the [11]
direction the FGGS are absent, while they are very large when the drive
is in the [10] direction.\cite{sohn} Another example of more recent interest
is the phenomenon of transverse critical current in superconductors with
pinning.\cite{transv} It has been found that for a 
VL a driven with a high
current there is a transverse critical current when an additional
small bias is applied
in the perpendicular direction.\cite{nori,nori2,md99} Furthermore, when the
transverse bias is increased it is possible to have a rich behavior with
a Devil's staircase in the transverse voltage.\cite{nori2}

In this paper we will study in detail the breaking of rotational invariance
in square JJA. In this case, 
the discrete lattice of Josephson junctions induces a
periodic egg-carton potential for the motion of vortices.\cite{jjpin} 
We will study here
how the voltage response depends on the angle of the current with respect to
the lattice directions of the square JJA. We will show that there are
preferred directions for vortex motion for which there is orientational
pinning. This leads to an anomalous transverse voltage when vortices are
driven in directions different from the symmetry directions. An analogous
effect of a transverse voltage due to the guided motion of vortices has been
observed in YBCO superconductors with twin boundaries.\cite{twins}
 Another related case is the intrinsic
breaking of rotational symmetry of d-wave superconductivity which causes an
angle-dependent transverse voltage for large currents.\cite{vicente}
Here we will show the differences and similarities of the square JJA with 
these problems.

The paper is organized as follows. In Sec. II we present the model equations
for the dynamics of the JJA, which will be solved in the numerical simulations.
In Sec. III we describe the experimental details of the JJA used in the
measurements. In Sec. IV we will present our results for both the
simulations and for the experiments. In particular, we will 
present experiments 
corresponding to the orientation for which the effect of a transverse
voltage is maximum. Finally in Sec. V we will compare our results with
other similar effects and discuss future directions of study.

\section{MODEL}

We study the dynamics of JJA using the resistively shunted junction (RSJ)\
model for the junctions of the square network.\cite{dyna,vorstroud,vorjose,nos}
In this case, the current
flowing in the junction between two superconducting islands in a JJA is
modeled as the sum of the Josephson supercurrent and the normal current: 
\begin{equation}
I_{\mu }({\bf n})=I_{0}\sin \theta _{\mu }({\bf n})+\frac{\Phi _{0}}{2\pi
cR_{N}}\frac{\partial \theta _{\mu }({\bf n})}{\partial t}+\eta _{\mu }({\bf %
n},t)
\end{equation}
where $I_{0}$ is the critical current of the junction between the sites $%
{\bf n}$ and ${\bf n}+{\bf \mu }$ in a square lattice [${\bf n}=(n_{x},n_{y})
$, ${\bf \mu }={\bf \hat{x}},{\bf \hat{y}}$], $R_{N}$ is the normal state
resistance and 
\begin{equation}
\theta _{\mu }({\bf n})=\theta ({\bf n}+{\bf \mu })-\theta ({\bf n})-A_{\mu
}({\bf n})=\Delta _{\mu }\theta ({\bf n})-A_{\mu }({\bf n})
\end{equation}
is the gauge invariant phase difference with 
\begin{equation}
A_{\mu }({\bf n})=\frac{2\pi }{\Phi _{0}}\int_{{\bf n}a}^{({\bf n}+{\bf \mu }%
)a}{\bf A}\cdot d{\bf l}.
\end{equation}
The thermal noise fluctuations $\eta _{\mu }$ have correlations 
\begin{equation}
\langle \eta _{\mu }({\bf n},t)\eta _{\mu ^{\prime }}({\bf n^{\prime }}%
,t^{\prime })\rangle =\frac{2kT}{R_{N}}\delta _{\mu ,\mu ^{\prime }}\delta _{%
{\bf n},{\bf n^{\prime }}}\delta (t-t^{\prime })
\end{equation}

In the presence of an external magnetic field $H$ we have 
\begin{eqnarray}
\Delta_{\mu}\times A_{\mu}({\bf n})&=&A_x({\bf n})-A_x({\bf n}+{\bf y})+ 
A_y({\bf n}+{\bf x})-A_y({\bf n})\nonumber\\
&=&2\pi f,
\end{eqnarray}
$f=H a^2/\Phi_0$ and $a$ is the array lattice spacing. We take periodic
boundary conditions (p.b.c) in both directions in the presence of an
external current ${\bf I}=(I_x,I_y)$ in arrays with $L\times L$ 
junctions.\cite{mio} The vector potential is taken as 
\begin{equation}
A_{\mu}({\bf n},t)=A_{\mu}^0({\bf n})-\alpha_{\mu}(t)
\end{equation}
where in the Landau gauge $A^0_x({\bf n})=-2\pi f n_y$, $A^0_y({\bf n})=0$
and $\alpha_{\mu}(t)$ allows for total voltage fluctuations under periodic
boundary conditions. In this gauge the p.b.c. for the phases are:\cite{md99,mio} 
\begin{eqnarray}
\theta(n_x+L,n_y)&=&\theta(n_x,n_y)  \nonumber \\
\theta(n_x,n_y+L)&=&\theta(n_x,n_y)-2\pi f Ln_x.
\end{eqnarray}
The condition of a total current flowing in the $x$ and $y$ directions: 
\begin{eqnarray}
I_x&=&\frac{1}{L^2}\left[\sum_{%
{\bf n}} I_0\sin\theta_x({\bf n})+\eta_x({\bf n},t)\right] + \frac{\hbar}{%
2eR_N} \frac{d\alpha_x}{dt}\;,  \nonumber \\
& & \\
I_y&=&\frac{1}{L^2}\left[\sum_{%
{\bf n}} I_0\sin\theta_y({\bf n})+\eta_y({\bf n},t)\right] + \frac{\hbar}{%
2eR_N} \frac{d\alpha_y}{dt}\;,\nonumber
\end{eqnarray}
determines the dynamics of $\alpha_\mu(t)$. \cite{mio} We also consider
local conservation of current, 
\begin{equation}
\Delta_\mu\cdot I_{\mu}({\bf n})=\sum_{\mu} I_{\mu}({\bf n})- I_{\mu}({\bf n}%
-{\bf \mu})=0.
\end{equation}
After Eqs.~(1,8,9) we obtain the following set of dynamical equations for
the phases,\cite{md99,mio} 
\begin{eqnarray}
\Delta_{\mu}^2\frac{\partial\theta({\bf n})}{\partial t}&=&-\Delta_{\mu}%
\cdot [S_{\mu}({\bf n})+\eta_{\mu}({\bf n},t)] \\
\frac{\partial\alpha_{\mu}}{\partial t}&=&I_{\mu} -\frac{1}{L^2}\sum_{{\bf n}%
} [S_{\mu}({\bf n})+\eta_{\mu}({\bf n},t)]
\end{eqnarray}
where 
\begin{equation}
S_{\mu}({\bf n})=\sin[\Delta_\mu\theta({\bf n})-A_{\mu}^0({\bf n})-
\alpha_{\mu}],
\end{equation}
we have normalized currents by $I_0$, time by $\tau_J=2\pi cR_{N}I_0/\Phi_0$%
, temperature by $I_0\Phi_0/2\pi k_B$, and the discrete Laplacian is 
\begin{eqnarray}
\Delta^2_\mu\theta({\bf n})&=&\theta({\bf n}+{\bf \hat x}) +\theta({\bf n}-%
{\bf \hat x})+\theta({\bf n}+{\bf \hat y}) +\theta({\bf n}-{\bf \hat y}%
)\nonumber\\
& &\;-4\,\theta({\bf n}).
\end{eqnarray}

The Langevin dynamical equations (10-11) are solved with a second order
Runge-Kutta-Helfand-Greenside algorithm with time step $\Delta t=0.1\tau_J$
and integration time $5000\tau_J$ after a transient of $2000\tau_J$. The
discrete Laplacian is inverted with a fast Fourier + tridiagonalization
algorithm as in Ref.\onlinecite{nos}. We calculate the time average of the
total voltage as 
\begin{eqnarray}
V_x&=&\langle v_x(t)\rangle= \langle d\alpha_x(t)/dt\rangle  \nonumber \\
V_y&=&\langle v_y(t)\rangle= \langle d\alpha_y(t)/dt\rangle
\end{eqnarray}
with voltages normalized by $R_ {N}I_0$. 

We study the JJA under a magnetic field corresponding to a single vortex in
the array, $f=1/L^2$, and system sizes of $L\times L$ junctions, with $L=32,64$%
. We apply a current $I$ at an angle $\phi$ with respect to the $[10]$
lattice direction, 
\begin{eqnarray}
I_x&=&I\cos\phi  \nonumber \\
I_y&=&I\sin\phi.
\end{eqnarray}
We define the longitudinal voltage as the voltage in the direction of the
applied current, 
\begin{equation}
V_l=V_x\cos\phi+V_y\sin\phi,
\end{equation}
and the transverse voltage 
\begin{equation}
V_t=-V_x\sin\phi+V_y\cos\phi.
\end{equation}
From the voltage response, we define the transverse angle as 
\begin{equation}
\tan\theta_t=V_t/V_l
\end{equation}
and the voltage angle as 
\begin{equation}
\tan\theta_v=V_y/V_x,
\end{equation}
i.e. $\theta_t=\theta_v-\phi$, see Fig.~1.
 
When the vortices move in the direction
perpendicular to the current, there is no transverse voltage, therefore $%
\theta_t=0$ and $\theta_v=\phi$.

\section{EXPERIMENTAL SETUP}

We measured Current--Voltage (IV) characteristics of square
proximity--effect Pb/Cu/Pb Josephson Arrays with the current applied in
different directions. 

The samples consist of 2500 \AA-thick  cross-shaped Pb islands on top of a
continuous 2500 \AA-thick copper film. Copper, and subsequently lead, were
thermally evaporated onto a silicon substrate within the same evaporator. 
An array of $1000\times1000$ lead islands were defined by photolithographic
patterning followed by an Ar ion etching. The cell parameter of the
resulting array was 10 $\mu$m with junctions 2 $\mu$m wide and a separation
of 1 $\mu$m.  A second photolithography step was used to define a
$1\times10$ mm$^2$ strip with current and voltage (longitudinal and
transversal) contacts. This mask was manually aligned in the [10] and [11]
directions for different samples.

The six-terminal measurements were made using a programable dc current
source and a two channel nanovoltmeter (HP 34420A). Each channel measuring
the longitudinal and transversal voltage respectively.  The arrays were
coooled up to 1.25 K in a pumped $^4$He cryostat shielded by $\mu$-metal. A
superconducting solenoid was used to null the remaining ambient magnetic
field ($\approx 15$ mG), and apply fields to the sample. The typical
periodic-in-field response in the resistance were observed extending in a
large number of periods, which was used to determine the frustration applied
to the sample. 
  
\section{NUMERICAL AND EXPERIMENTAL RESULTS}

\subsection{Breaking of rotational invariance.}

The square lattice has two directions of maximum symmetry: the [10] and the
[11] directions (and the ones obtained from them by $\pi/2$ rotations),
which correspond to the directions of reflection symmetry. When the current
bias is in the [10] direction, the angle of the current is $\phi=0$, and we
call it a ``parallel" bias. When the current bias is in the [11] direction,
the angle of the current is $\phi=\pi/4=45^o$, and we call it a ``diagonal"
bias.


In the case of the parallel bias we find that the transverse voltage is zero
(in agreement with the reflection symmetry). This corresponds to vortex
motion in the direction perpendicular to the current ($\theta_t=0$). 
In the IV curve for the longitudinal voltage we find a
critical current corresponding to the single vortex depinning $I_c^{[10]}=0.1
$,\cite{jjpin} as shown in Fig.~2(a). Above $I_c$ it is possible to distinguish three
regimes: \cite{vorstroud,vorjose} ({\bf A}) a single vortex regime for $%
0.1<I<0.85$, where the IV is quasi-linear and dissipation is caused by
vortex motion; ({\bf B}) an intermediate crossover regime for $0.85<I<1.0$
and ({\bf C}) a resistive regime for $I>1.0$ where dissipation is caused by
the Ohmic shunt resistance and the IV curve is linear. The vortex regime 
{\bf A} can be described by the dynamics of a single collective degree of
freedom: an overdamped particle moving in the periodic ``egg-carton"
potential,\cite{jjpin,vorstroud} although with a non-linear viscosity.\cite
{vorjose} The resistive regime {\bf C} is also very simple, it can be
represented by the behavior of a single junction at large currents, $%
\theta_\mu({\bf n},t)\approx\frac{2eR_N}{\hbar}I_\mu t +\delta_\mu({\bf n},t)
$. The crossover regime {\bf B} is characterized by a complex collective
dynamics with an interplay of the vortex degree of freedom with spin waves
excitations, and at finite temperatures there is also a steep increase of
vortex-antivortex excitations in this regime.

In the case of the diagonal bias, we obtain similar results as in the
parallel bias case. The transverse voltage is zero and therefore the vortex
moves perpendicular to the current ($\theta_t=0$). The IV curve for $V_l$
has a critical current of $I_c^{[11]}=\sqrt{2}I_c^{[10]}=0.1414$. The onset
of the resistive regime {\bf C} is also multiplied by a factor of $\sqrt{2}$%
, while the crossover regime {\bf B} starts at nearly the same current as in
the parallel bias case, see Fig.~2(b).


For orientations different than the symmetry directions, we always find a
finite transverse voltage. In order to see this, we study the voltage
response when varying the orientation $\phi$ of the drive while keeping
fixed the amplitude $I$ of the current. In Fig.~3(a) we plot the transverse
angle $\theta _{t}=\arctan ({V_{t}}/{V_{l}})$ as a function of the angle of
the current $\phi$. We find that $\theta_t$ vanishes only in the maximum
symmetry directions corresponding to angles $\phi =0,\pm 45^{o},\pm
90^{o},\dots$, as discussed before. Furthermore, we see that for
orientations near $\phi=0$, the transverse angle basically follows the
current angle: $\theta _{t}\approx -\phi $. This is an indication that
vortex motion is pinned in the lattice direction [10], since $V_{y}\approx 0$%
, meaning that the voltage angle is $\theta _{v}\approx 0$. 
Whenever the voltage response is
insensitive to small changes in the orientation of the current, we 
will call this phenomenon {\it orientational pinning}. On the other
hand, near $\phi=45^{o}$ the transverse angle changes rapidly. We show in
Fig.~3(b-c) the behavior of $\theta_{t}$ for different currents and
temperatures. The $\theta_{t}$ vs. $\phi $ curves become smoother around $%
\phi =45^{o}$ for increasing current as well as for increasing $T$. At the
same time, the magnitude of the transverse angle decreases when increasing
the current for $I\gg 1$, or when increasing $T$.

A more direct evidence of the breaking of rotational symmetry 
can be seen in the parametric curves of $V_{y}(\phi )$
vs. $V_{x}(\phi )$. In Fig.~4 we plot the values of the voltages $V_{y}$ and 
$V_{x}$ when varying the orientational angle $\phi $ for different values of
the current amplitude $I$ and the temperature. In the case of rotational
symmetry we should have a perfect circle. In the set of plots of Fig.4(a-c),
the current amplitude is fixed and the temperature is varied. In Fig.4(a) we
have $I=0.2$, near the onset of single vortex motion in the regime {\bf A}.
In this case most of the points are either on the axis $V_{x}=0$ or on the
axis $V_{y}=0$, indicating strong orientational pinning in the lattice
directions [10] or [01]. When increasing $T$ the orientational pinning
decreases and the length of the ``horns'' in the $x$ and $y$ axis decreases.
Fig.4(b) corresponds to $I=0.6$ , near the end of regime {\bf A } when the
vortex is moving fast. In this case the horns have disappeared and
orientational pinning is lost. However, the breaking of rotational symmetry
is still present in the star-shaped curves that we find at low $T$. The dip
at $45^0$ in the stars are because in this direction the voltages are
minimum, since the critical current is maximum in this case, $I_c^{[11]}=0.1%
\sqrt{2}$. When increasing the temperature, the stars tend to the circular
shape of rotational invariance. The parametric curves in the crossover
regime {\bf B} also have star shaped behavior at low $T$ which tends to
circles when increasing $T$. Above the onset of the resistive regime {\bf C}
the ``horned'' curves reappear [Fig.~4(c)]. 
In this case the orientational pinning
corresponds to the locking of ohmic dissipation in the junctions in one of
the lattice directions, either [10] or [01]. Once again, when increasing $T$ the
horned structure shrinks, and the curves evolve continuously from square shapes
to circular shapes.

The variation with current of the rotational parametric curves for a fixed
temperature is shown in Figs.4(d-f). At a low temperature, $T=0.05$, we
clearly see the horned structure of the curves for almost all the currents
and even for large currents the circular curves have ``horns'', see
Fig.4(d). At an intermediate temperature, $T=0.2$, there are still some
signatures of the orientational pinning [Fig.~4(e)], while for $T=1.2$ all the curves
are smooth and rounded with a slightly square shape [Fig.~4(f)].

\subsection{Orientational pinning near the [10] direction.}

The orientational pinning characterizes the breaking of rotational symmetry
in square arrays at low temperatures, as we have seen in the previous Section. 
When the
JJA is driven in any of the ``parallel'' directions, $\phi=0^o ,90^o ,180^o
,270^o$ both vortex motion and dissipation are pinned along these
directions. When the current is rotated a small angle, for example from the
[10] direction, the dissipation remains pinned along the [10] direction ($V_x%
{\not=}0$, $V_y=0$). This effect causes a finite transverse voltage,
measured with respect to the direction of the current, $V_t\approx
-V_x\sin\phi$ and a transverse angle $\theta_t=-\phi$. The orientational
pinning is lost at a given critical angle $\phi_c$ which depends on
temperature and current. As we can see in Fig.~3(a), for very low $T$ and
for certain values of the current, the critical angle can reach values very
close to $45^o$. This leads to a large transverse voltage in arrays driven
near the [11] direction as we will discuss later in Sec.~IV.C.

Let us analyze now the behavior of the critical angle  for
orientational depinning, $\phi _{c}$. 
This can be studied by looking at the angle of the
voltage with respect to the [10] direction, $\theta _{v}=\arctan
(V_{y}/V_{x})$. For $\phi <\phi _{c}$ we have orientational pinning and
therefore $\theta _{v}=0$, while for $\phi >\phi _{c}$ we have $\theta _{v}%
{\not=}0$. Therefore, the onset of a finite $\theta _{v}$ defines the
critical angle $\phi _{c}$. In Fig.~5 we plot $\theta _{v}$ as a function of 
$\phi $ for different currents and temperatures. There are two cases of
interest: for a current in the single vortex regime {\bf A} [Fig.~5(a)] and
for a current in the resistive regime {\bf C} [Fig.~5(b)]. We find that in
both cases there is clearly a finite $\phi _{c}$ which decreases with
temperature. In the case of the single vortex regime {\bf A}, we see that $%
\phi _{c}$ tends to vanish at a crossover temperature corresponding to the
energy scale for vortex depinning,\cite{jjpin}
 $T_{pin}\sim \Delta E_{pin}\approx 0.2$.
 The notion of ``transverse critical current'' $I_{c,tr}$ has
been introduced recently by Giamarchi and Le Doussal for driven vortex
lattices in random pinning. \cite{transv} When a vortex structure is moving fast, it can
still be pinned in the transverse direction. After applying a current in the
direction perpendicular to the drive, a finite transverse critical current
may exist at $T=0$. In the case of periodic pinning, a finite $I_{c,tr}$ was
also proposed due to conmensurability effects.\cite{transv} In particular, in
the periodic ``egg-carton'' pinning of Josephson junction arrays it was
found in Ref.\onlinecite{md99} that $I_{c,tr}$ is finite in a wide range of
temperatures for a system driven in the [10] direction. Here we see that due
to the orientational pinning, a transverse critical current is finite only
when the JJA is driven either in the [10] or [01] directions. In any other
case it will be zero. Moreover, the critical angle in the regime {\bf A} can
be interpreted as corresponding to a transverse critical current $%
I_{c,tr}=I\sin {\phi _{c}}$ for a vortex driven by a longitudinal current $%
I_{l}=I\cos {\phi _{c}}$.

In Fig.~5(b) we see that for a current in the resistive regime {\bf C} there
is also a critical angle for the onset of transverse 
ohmic dissipation. In this case $%
\phi_c$ tends to vanish at a higher temperature above the
Kosterlitz-Thouless transition, $T_{KT}\approx0.9$.

The other lattice symmetry direction of interest corresponds to the case of
``diagonal'' bias, i.e. the [11] directions of $\phi=45^o,135^o,225^o,315^o$%
. In this case, we have shown previously that the transverse voltage also
vanishes. In agreement with this, we see both in Fig.~5(a) and Fig.~5(b) that
all the curves cross in the point $(\phi,\theta_v)=(45^o,45^o)$, which
corresponds to $\theta_t=\theta_v-\phi=0$. However, there is no
orientational pinning in the [11] direction. On the contrary, in this
direction any small deviation in the orientation of the current can cause a
fast increase of the transverse voltage. In order to show this effect, we
study the ``diagonal voltage angle'', which is measured with respect to $45^o
$: $\theta_v^{\prime}=\arctan{(V_y-V_x)/(V_y+V_x)}$ $= \theta_v-\pi/4$, as a
function of the deviation from the [11] direction, $\Delta\phi=\phi-\pi/4$,
see Fig.~6. We find that, in contrast with the case of Fig.~5, a small
deviation from the symmetry direction leads to a large change in the voltage
angle for any current and temperature. This shows that the [11] direction is
highly unstable against small changes in the orientation of the current,
leading to an anomalously large transverse voltage.

\subsection{Transverse voltage near the [11] direction.}

In Fig.~7(a) we show our experimental voltage-current characteristics for an
array of $100\times 1000$ junctions at a low temperature $T=1.25K$ and at a
low magnetic field. The current is applied nominally in the [11] direction,
but a small misalignment is possible in the setup of electrical contacts,
therefore $\phi =45^{o}\pm 5^{o}$. We see that for low currents there is a
very large value of the transverse voltage $V_{t}$, which is nearly of the same
magnitude as the longitudinal voltage $V_{l}$. The transverse voltage is
maximum at a characteristic current $I_{m}$. Above $I_{m}$, $V_{t}$
decreases with increasing current while $V_{l}$ increases. It is remarkable
that these results are very different from the IV curve of Fig.~2, where $%
V_{t}=0$ at $\phi =45^{o}$. However, if we assume a misalignment of a few
degrees with respect to the [11] direction we can reproduce the experimental
results. In Fig.~7(b) we show the IV curves obtained numerically for $\phi
=40^{o}$ and $T=0.02$. We see that for low currents $V_{t}$ is close to $%
V_{l}$: $V_{t}\lesssim V_{l}$, similar to the experiment, and later $V_{t}$
has a maximum at a current $I_{m}\approx 1/\cos \phi \approx 1.3$. This
corresponds to the current for which the junctions in the $x$-direction
become critical ($I_{x}=1$). The range of currents we can measure
experimentally is limited to the regimes {\bf B} and {\bf C}, since we can
not fully access the regime {\bf A} of single vortex motion due to the small
voltages involved in this case. Of course, in the simulations we can study
the full range of currents, which is shown in Fig.~7(c). Here we see that
near the vortex depinning current the transverse voltage is also very close
to $V_l$ in a small range of currents, then when increasing $I$ they separate
first inside the regime {\bf A}, and later in the regime {\bf B} the
transverse voltage approaches the longitudinal voltage again.

As we saw in Fig.~3 the highest transverse voltage can be obtained for
orientations near $\phi=45^o$. Therefore a slight misalignment of the array
from the [11] direction is useful for studying both experimentally and
numerically the behavior of the transverse voltage as a function of current
and temperature.

In Fig.~8 we show the dependence of $\theta_t$ with current for different
temperatures. The experimental results are shown in Fig.8(a) where we find
that $\theta_t$ first increases with current, it reaches a maximum  value $%
\theta_t^{max}(T) < 45^o$ and then for large currents $\theta_t$ tends to
zero. In Fig.~8(b) we show that the simulations with the RSJ model with $%
\phi=40^o$ reproduce this behavior. Here we see that the maximum of the
transverse angle is reached inside the regime {\bf B} well before the onset
of the resistive regime. We also find that, when going deep into the regime 
{\bf C}, $\theta_t$ decreases with current: $\theta_t \rightarrow 0$ for $%
I\gg I_m$. In Fig.~8(c) we show that a similar behavior is obtained for other
values of $\phi$ close to $45^o$. We also observe here the full range of
currents. We see that at the critical current the transverse angle $\theta_t$
first has a maximum, then it decreases rapidly in a small range of currents,
after which, for most of the regime {\bf A}, $\theta_t$ is increasing with $I
$ before reaching a second maximum value in the crossover regime {\bf B}.
This shows that there are two regimes where the effect of anomalous
transverse voltage is maximum: near the vortex depinning current, due to
orientational pinning of vortex motion; and near the Josephson junction
critical current due to the orientational ``pinning'' of ohmic dissipation.
Regretfully, we can not measure the small voltages of the low current
regime, therefore we were not able to observe experimentally the first
maximum of $\theta_t$.

In Fig.~9 we analyze the behavior of the transverse angle as a function of
temperature. We plot the value of $\theta_t$ for a current near the maximum
value of the transverse voltage at low $T$. We observe experimentally that $%
\theta_t$ decreases with temperature and in particular it has a sharp decrease
at $T\approx 1.5K$, as we show in Fig.~9(a). On the other hand, in the
simulation results we see a smooth decrease of $\theta_t$ with temperature
[Fig.~9(b)]. The transverse angle becomes small at the depinning crossover
temperature $T_{pin}\approx 0.2$. The fact that our experimental results
show a sharper decrease with temperature than the simulations is possibly
due to vortex collective effects. The simulation results presented here are
focused in the motion of a single vortex in the periodic pinning of a square
JJA. The vortex collective effects, which have to be studied for fields $%
f>1/L^2$, will be discussed elsewhere. \cite{futuro}

\section{DISCUSSION}

In the egg-carton potential of a square JJA  there are pinning barriers for
vortex motion in all the directions. The direction with the lowest pinning
barrier is the [10] direction. Therefore the strong orientational pinning we
find here is in the direction of the lowest pinning for motion, i.e. the
direction of easy flow for vortices. The presence of a strong orientational
pinning leads to a large transverse voltage when the systems is driven away
from the favorable direction, to the existence of a critical angle and to a
transverse critical current. On the other hand, the [11] direction is the
direction of the largest barrier for vortex motion in the egg-carton
potential. In this case, the behavior is highly unstable against small
variations in the angle of the drive, leading to a rapid change from zero
transverse voltage to a large transverse voltage within a few degrees. This
explains the transverse voltage observed in our experimental measurements in
JJA\ driven near the diagonal [11] orientation.

An analogous effect of orientational pinning has also been seen in
experiments on YBCO
superconductors with twin boundaries. \cite{twins} In this case, due to the
correlated nature of the disorder, the direction for easy flow is the
direction of the twins. A similar effect of horns in the parametric voltage
curves are therefore observed in the direction corresponding to the twins.
Also transport measurements when the sample is driven at an angle with
respect to the twin show a large transverse voltage.

It is interesting to compare with the angle-dependent transverse voltage
calculated for d-wave superconductors.\cite{vicente} Also in this case, the
transverse voltage vanishes only in the [10] and [11] directions.  However
the $\theta_{t}$ vs. $\phi $ curves are smooth in this case, since
$\tan\theta_t\propto \sin4\phi$.\cite{vicente} This is because there is no
pinning and the transverse voltage is caused only by the intrinsic nature of
the d-wave ground state. On the other hand, the breaking of rotational
symmetry studied here is induced by the pinning potential, and it results in
non-smooth responses like ``horned'' parametric voltage curves, critical
angles, transverse critical currents, etc.

In superconductors with a square array of pinning centers, typically the
pins are of circular shape and the size of the pins is much smaller than the
distance between pinning sites.\cite{nori0,nori,nori2} In this case, the
pinning barriers that vortices find for motion are the same in many
directions. Therefore it is possible to have orientational pinning in many
of the square lattice symmetry directions. This explains the rich structure
of a Devil's staircase observed recently in the simulations of Reichhardt
and Nori, where each plateau corresponds to orientational pinning in each of
the several possible directions for orientational pinning.  This interesting
behavior is not possible in JJA, however, since the egg-carton pinning
potential corresponds to the situation of square-shaped pinning centers with
the pin size equal to the interpin distance.  In this case the only possible
directions for orientational pinning are the [10] and [01], as we have seen
here.

It is worth noting that many experiments in JJA in the past have been done
in samples with a diagonal bias.  For example, van Wees {\it et
al.}\cite{wees} have observed the existence of a transverse voltage in their
measurements, which was unexplained. A small misalignment of their direction
of the current drive could easily explain their results, since we have
learned here that the [11]-direction is unstable against small changes in
the angle of the bias. Also Chen {\it et al.}\cite{chen} have reported a
transverse angle in measurements in JJA driven in the diagonal direction. In
their case the effect has a strong component that is antisymmetric against a
change in the direction of the magnetic field, which means that they have a
Hall effect possibly due to quantum effects. However, they report that their
transverse voltage had also a component which was even with the magnetic
field (which was discounted in their computation of the Hall angle). This
particular spurious contribution can also be attributed to a small
misalignment of the direction of the bias.  From this we conclude that in
order to study the Hall effect in JJA the most convenient choice would be a
current bias in the [10] direction where the effect of transverse voltages
at small deviations is minimum.

When this work was upon completion, new studies of the effect of the
orientation of the bias in driven square JJA have appeared. Fisher, Stroud
and Janin \cite{stroud99} have studied some of the effects of the direction
of current in a fully frustrated JJA ($f=1/2$) at $T=0$. In their case a
transverse critical critical current and the dynamics as a function of $I_x$
and $I_y$ has been described. Their results are in part complementary to our
work with a single vortex ($f=1/L^2$).  Yoon, Choi and Kim \cite{choi} find
differences in the IV characteristics of JJA at $f=0$ when comparing the
parallel current bias with the diagonal current bias. Their results are in
agreement with our Fig.~2 results.

In this paper we have considered  the dynamics of a single vortex in a
square JJA.  We were able to characterize in detail the orientational
pinning and breaking of rotational symmetry in this case. Furthermore, with
the results of the RSJ numerical calculation we were able reproduce and
interpret most of our experimental measurements for a quasi-diagonal bias. 
It remains for the future to study the behavior of a driven vortex lattice
(VL) when the current is rotated, since the VL has also its own periodicity
and symmetry directions.  As we saw recently in Ref.~\onlinecite{md99}, a
moving vortex lattice in a JJA shows different dynamical phases as a
function of temperature and current.  Therefore we expect that the
characteristics of the breaking of rotational invariance, orientational
pinning and transverse voltages will depend on the dynamical phase under
consideration. \cite{futuro}

\acknowledgements
We acknowledge financial 
support from CONICET, CNEA, ANPCyT, Secyt-Cuyo, and Fundaci\'{o}n Antorchas.
On of us (P.\ M.) thanks  the Swiss National Science Foundation for support.

\begin{figure}[tbp]
\caption{Parametric curves $V_y(\phi)$ vs. $V_x(\phi)$.
(a) $I=0.2$, for (going
outwards from the center) $T=0$, $T=0.01$ and $T=0.05$; 
(b) $I=0.6$ for $T=0.3$, $T=0.5$, $T=0.6$, $T=1.0$ and $T=1.4$;
(c) $I=1.2$ for $T=0$, $T=0.05$, $T=0.1$, $T=0.3$ and $T=1.0$. 
(d) $T=0.05$  for (going
outwards from the center) $I=1.0$, $I=1.2$, $I=1.4$ and $I=1.6$;
(e) $T=0.2$ for $I=1.0$, $I=1.2$, $I=1.4$ and $I=1.6$; 
(f) $T=1.0$ for $I=0.2$, $I=0.4$, $I=0.6$ and $I=0.8$.
}
\label{fig3}
\end{figure}

\begin{figure}[tbp]
\caption{Orientational pinning near $\phi=0$.
Voltage angle $\theta_v=\arctan(V_y/V_x)$ vs. $\phi$ for different
temperatures and (a) $I=0.6$ (b) $I=1.2$. }
\label{fig7}
\end{figure}

\begin{figure}[tbp]
\caption{Voltage angle with respect to the $[11]$ direction $%
\theta_v^{\prime}=\arctan{(V_y-V_x)/(V_y+V_x)}$ $= \theta_v-\pi/4$ vs. $%
\phi-\pi/4$ for different temperatures and (a) $I=0.6$ (b) $I=1.2$.}
\label{fig8}
\end{figure}

\begin{figure}[tbp]
\caption{(a) Experimental results of longitudinal and transverse voltage for
a current near the [11] direction, $\phi=45^o\pm5$ at $T=1.25$ K. (b)
Longitudinal and transverse voltage obtained numerically for $\phi=40^o$ at $%
T=0.02\, \hbar I_0/2ek_B$. (c) Idem (b) for an extended current range.}
\label{fig4}
\end{figure}

\begin{figure}[tbp]
\caption{(a) Experimental results of transverse angle $\theta_t$ vs. current
for $\phi=45^o\pm5$ at different temperatures. (b) Numerical results of $%
\theta_t$ vs. $I$ for $\phi=40^o$. (c) $\theta_t$ vs $I$ 
at $T=0.02$ for different $\phi$ and full range of currents.
Experimental temperatures are in Kelvin, simulation temperatures
are in units of $\hbar I_0/2ek_B$.}
\label{fig5}
\end{figure}

\begin{figure}[tbp]
\caption{(a) Experimental results of transverse angle vs. $T$ for $%
\phi=45^o\pm5$ and $I=300\,\mu$A. (b) Numerical results of $\theta_t$ vs. $T$ for $%
\phi=40^o$ and $I=0.85$.}
\label{fig6}
\end{figure}
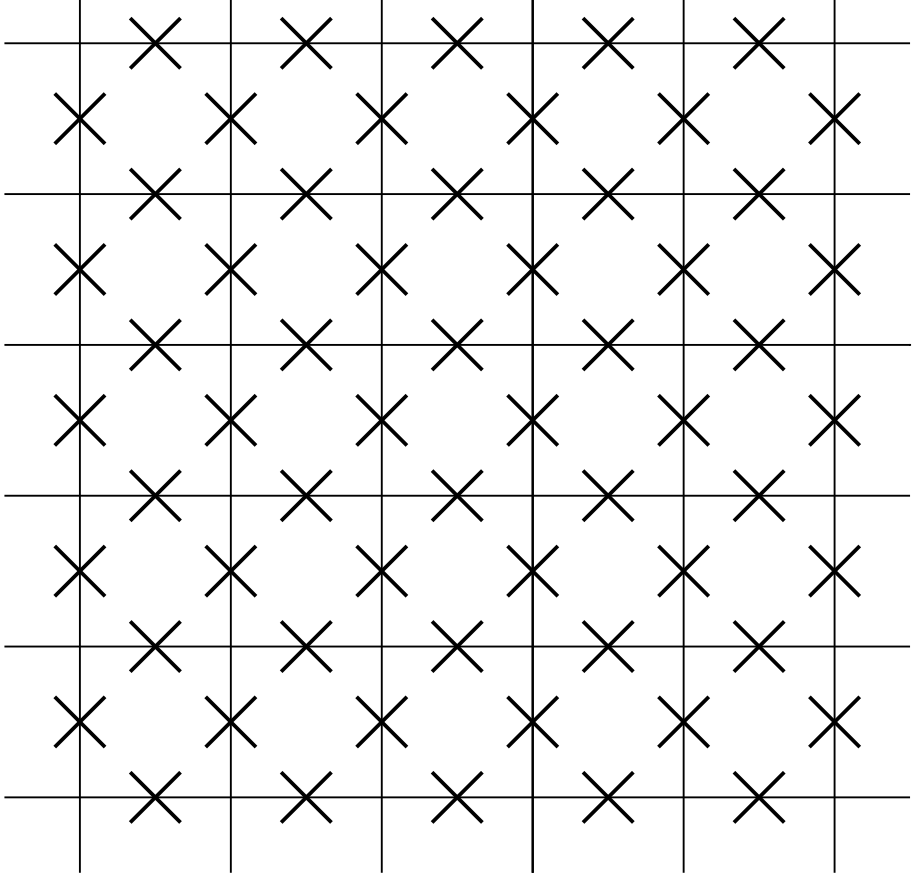
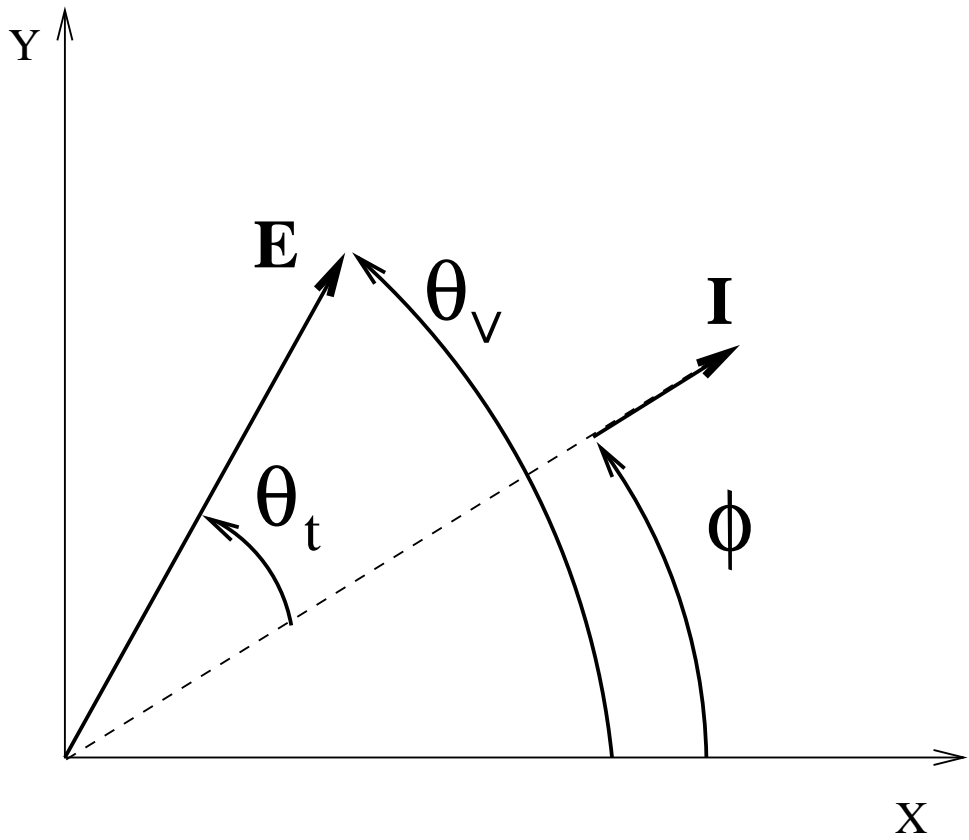
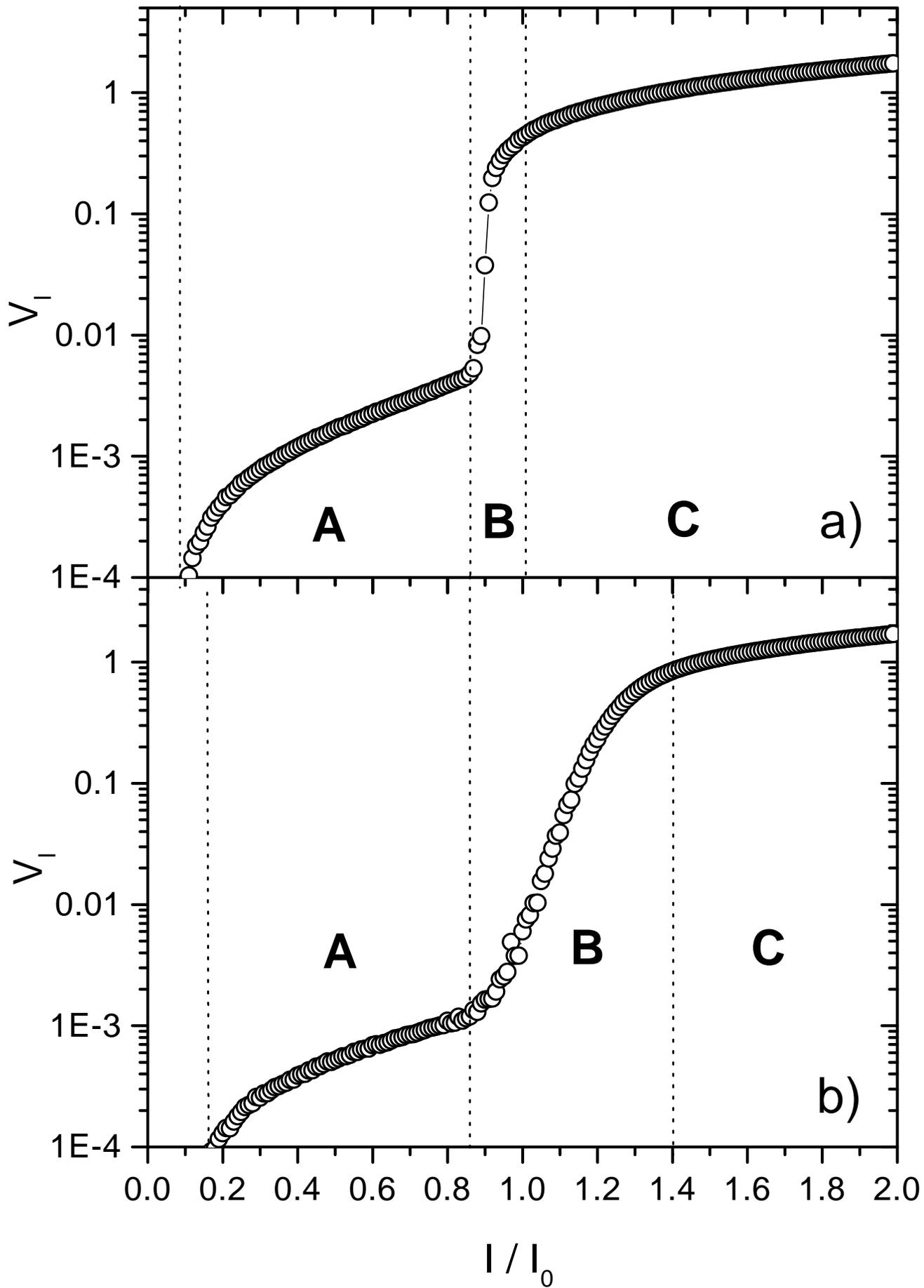
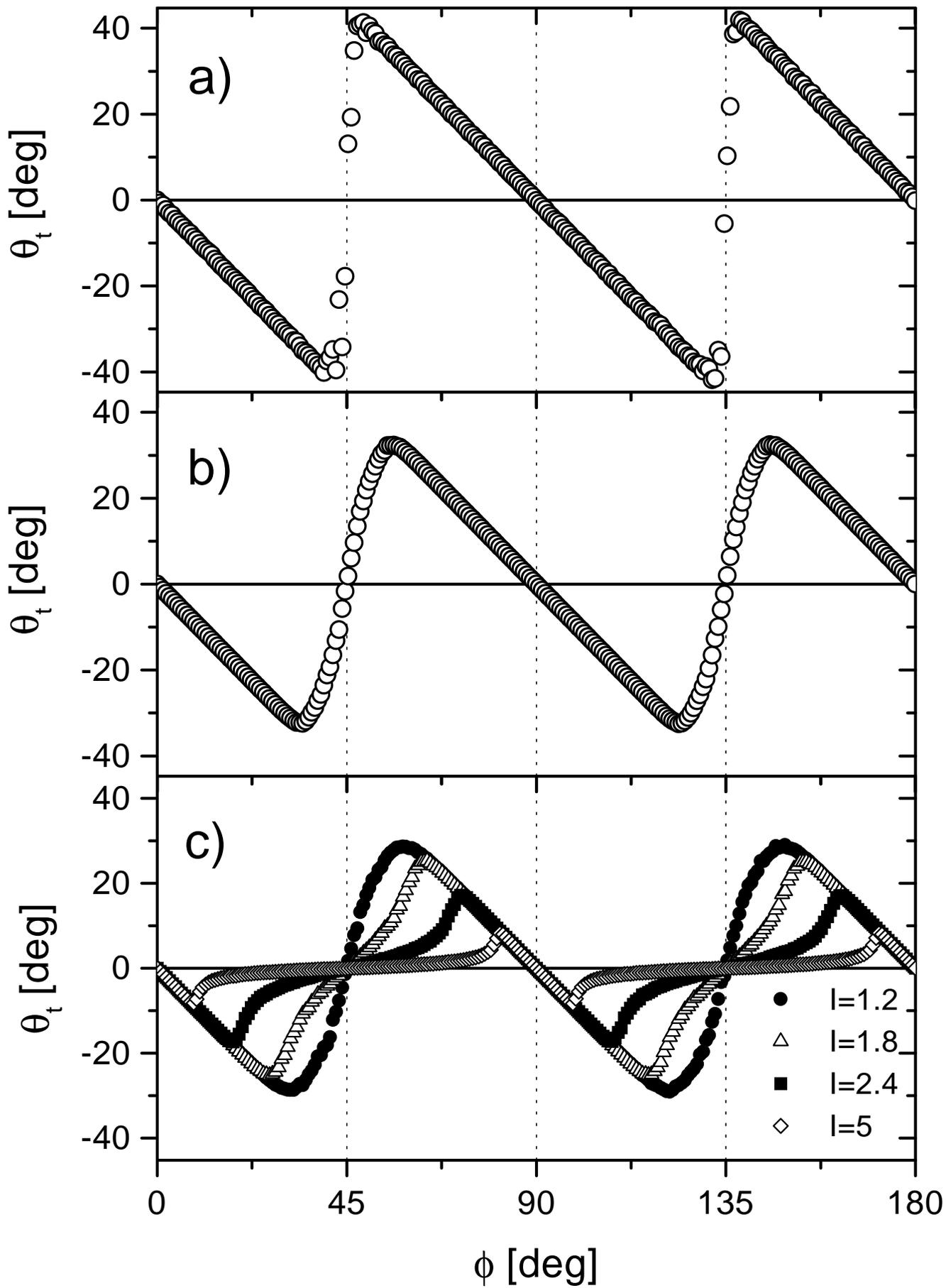
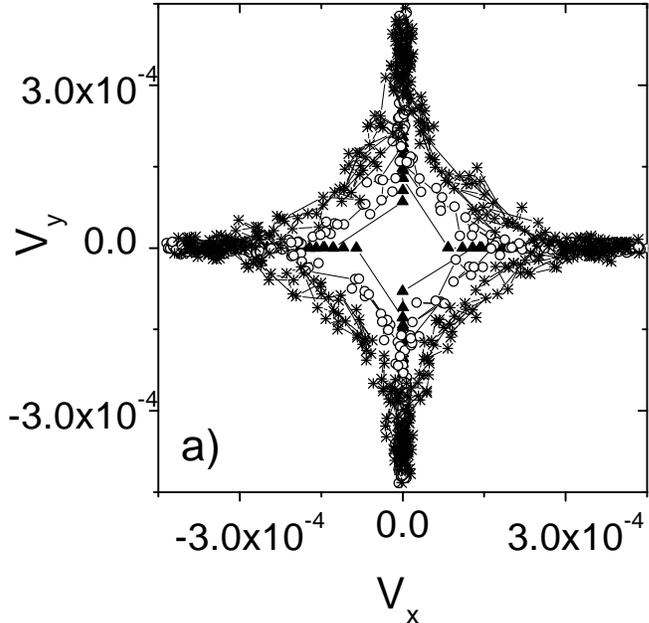
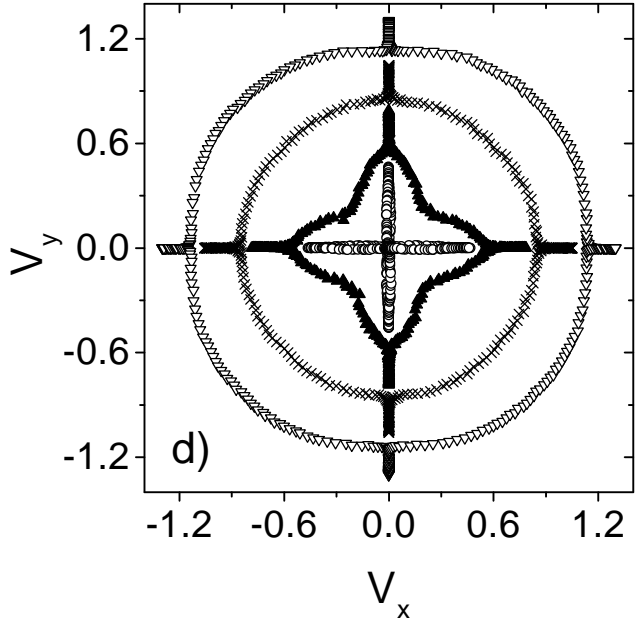
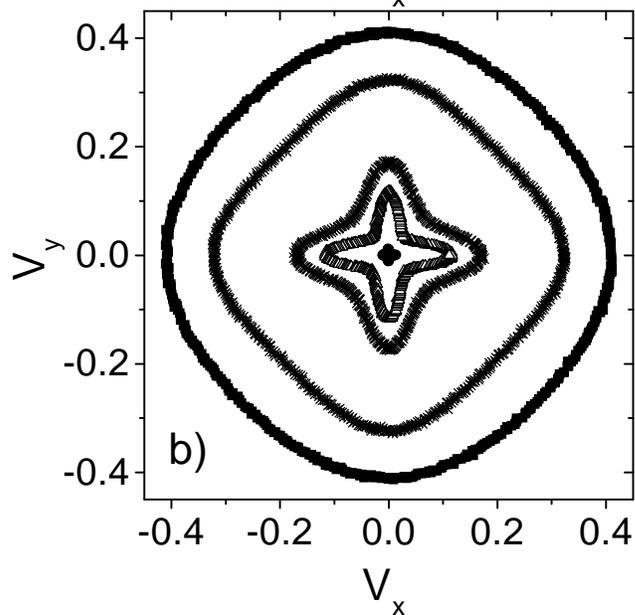
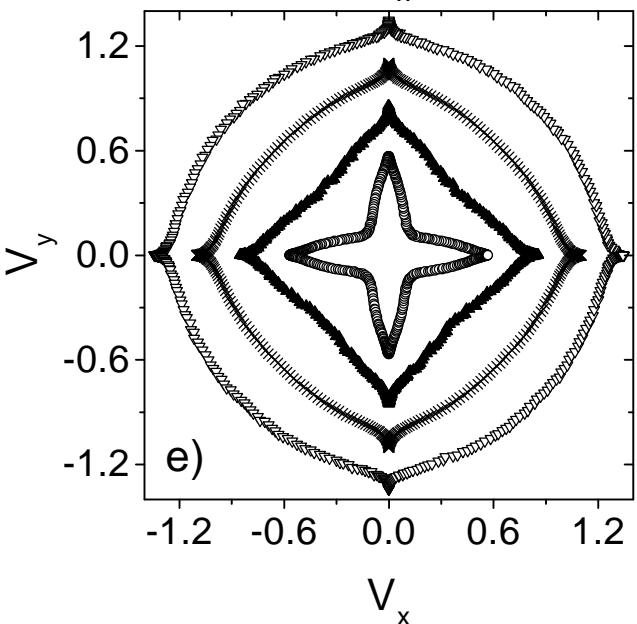
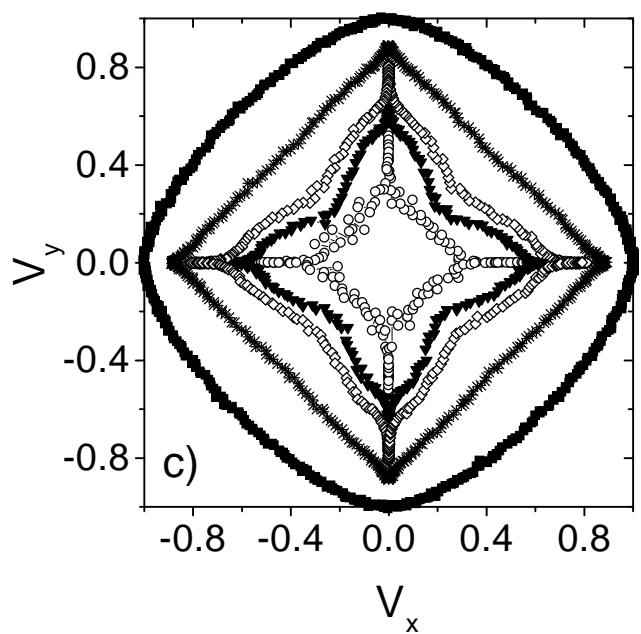
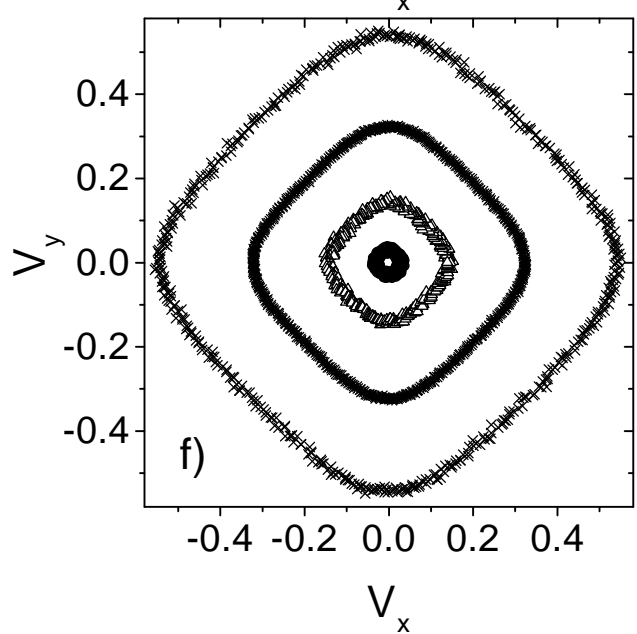
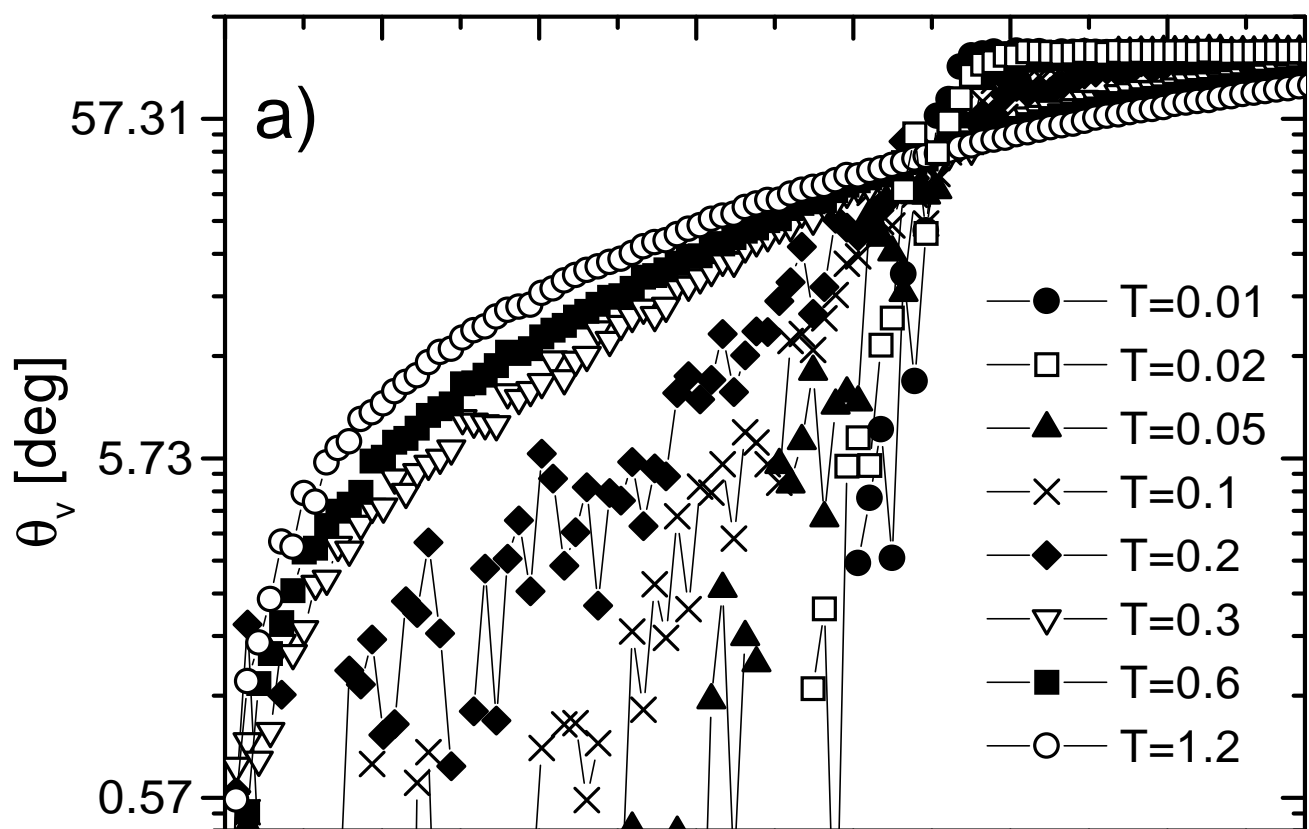
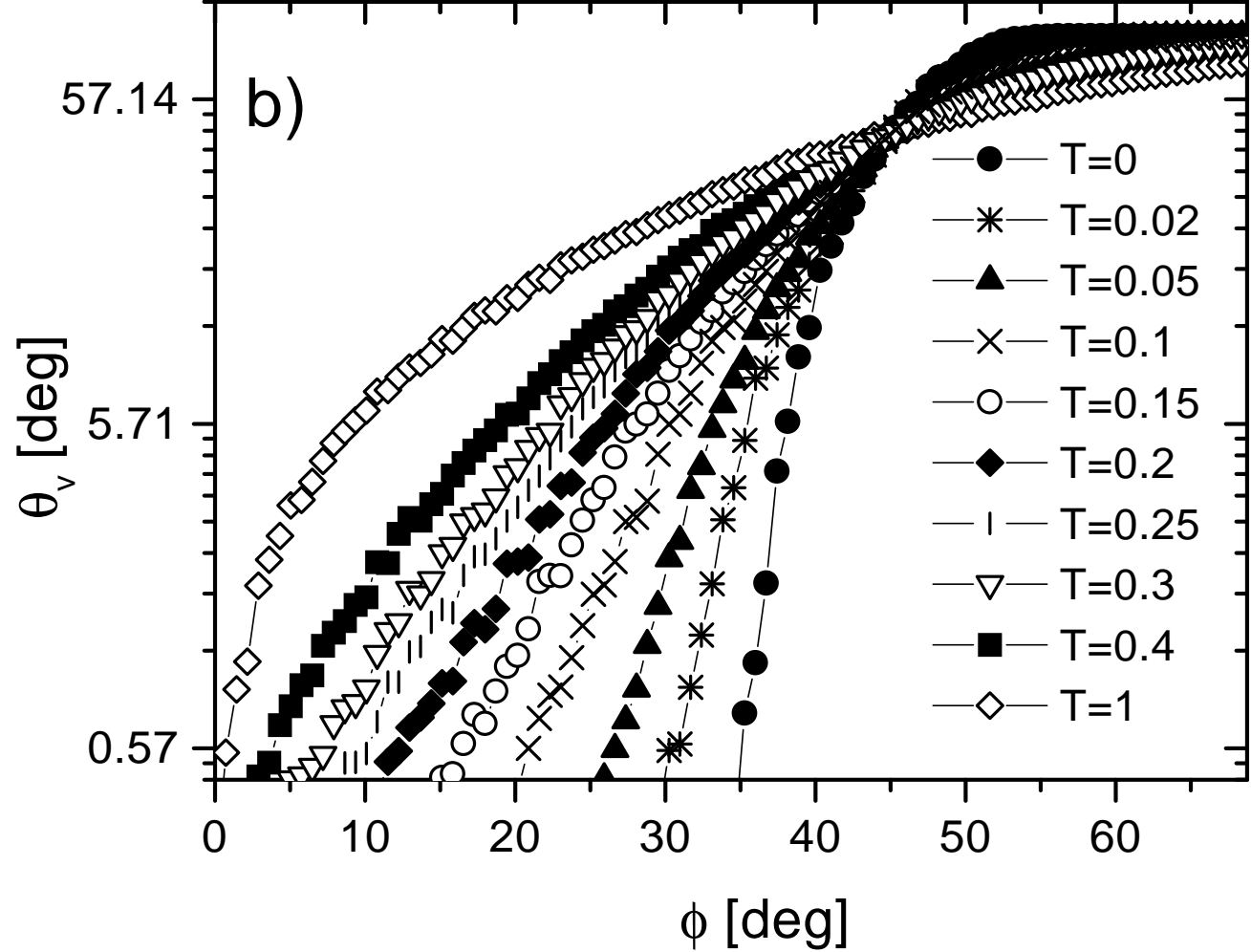
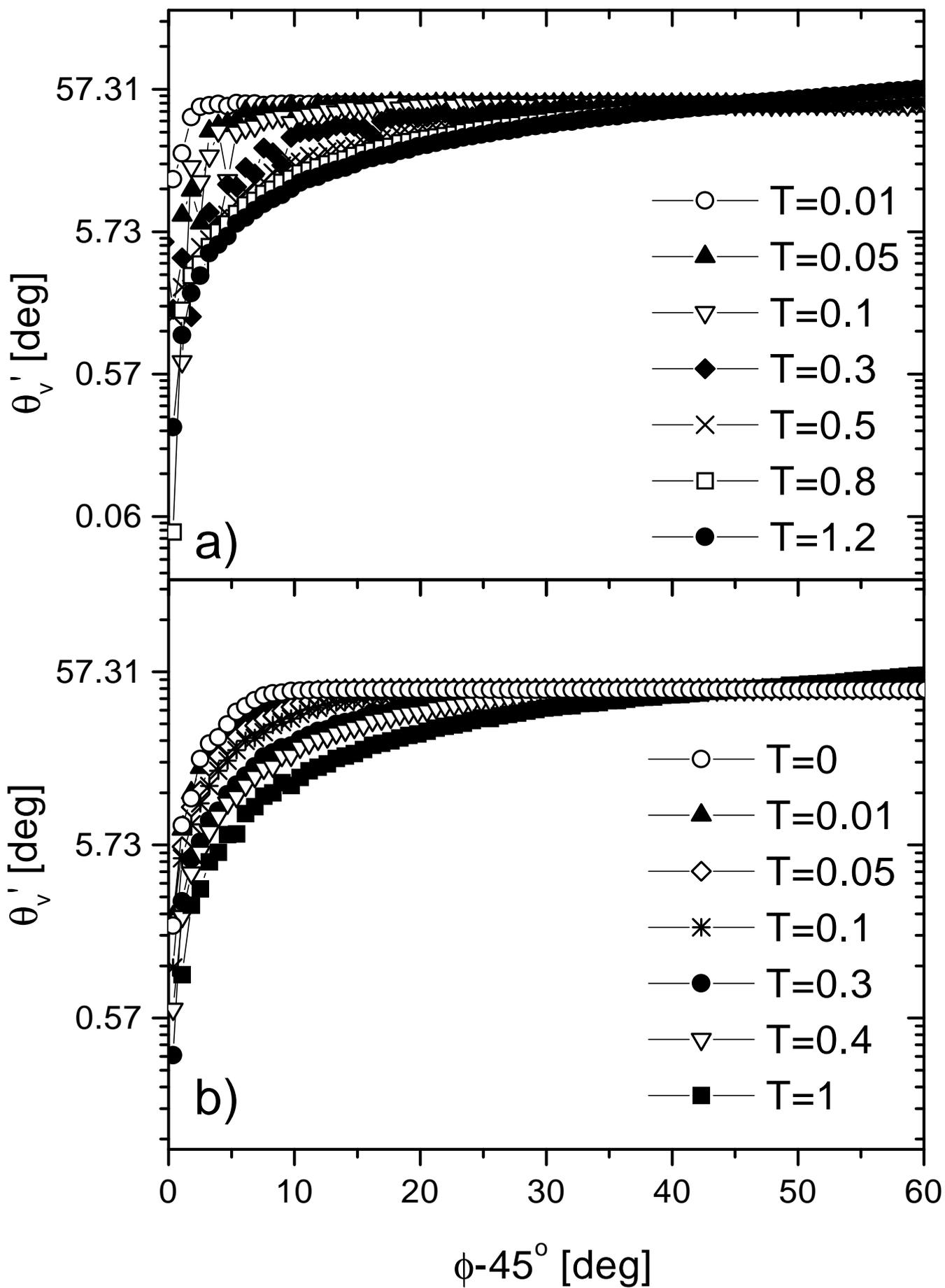
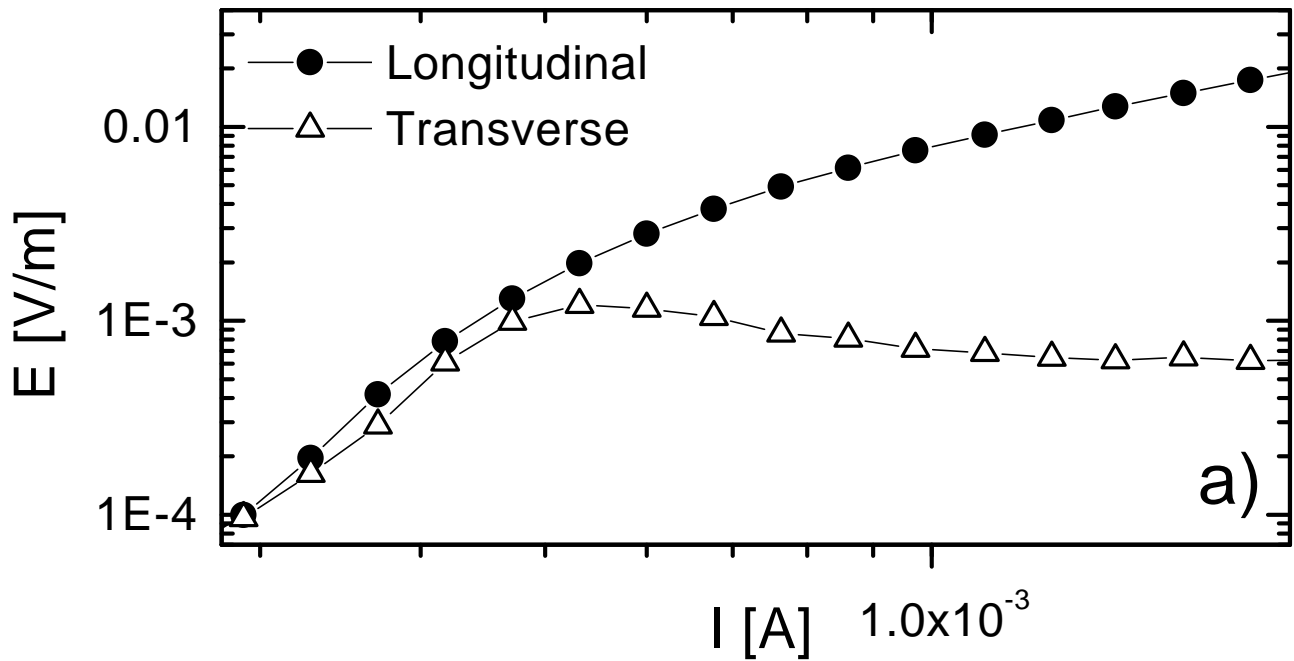
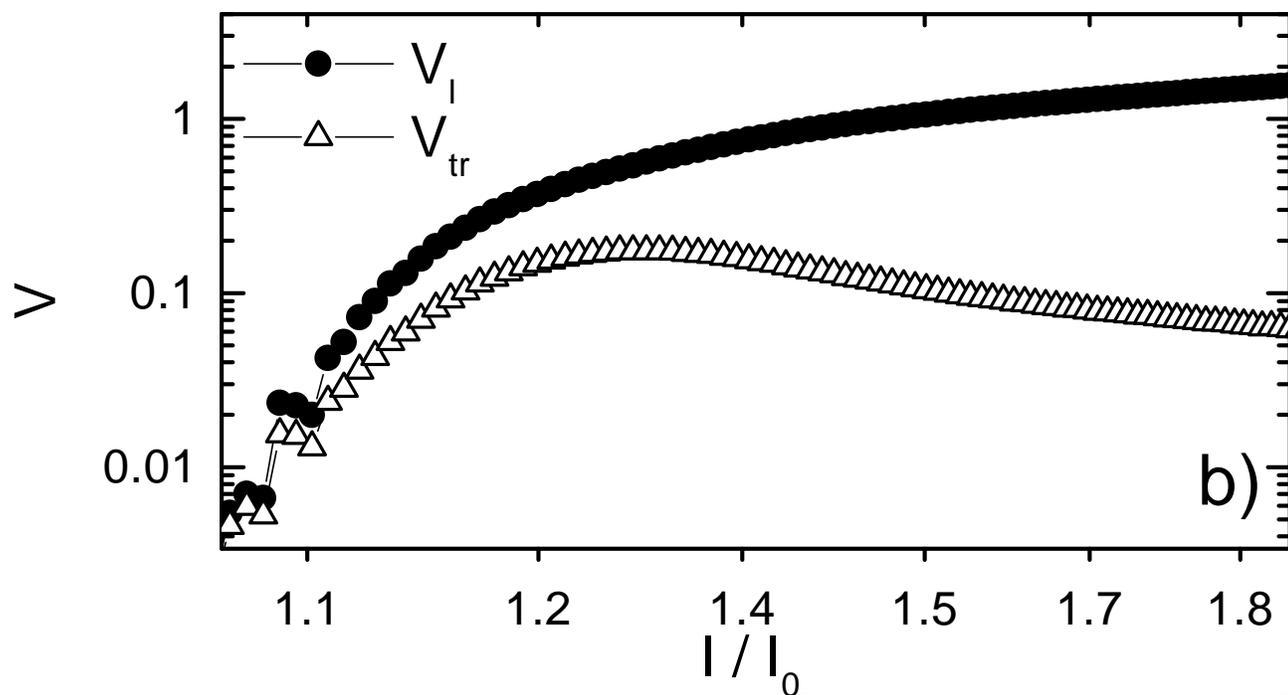
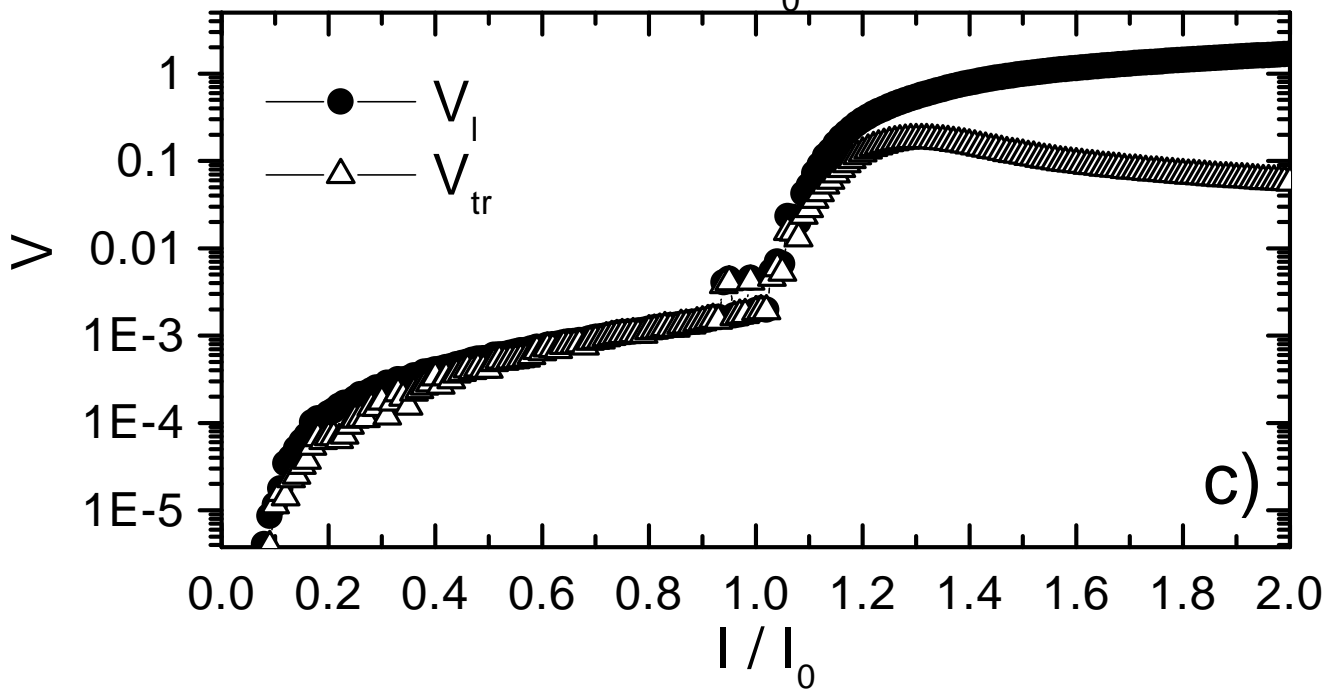
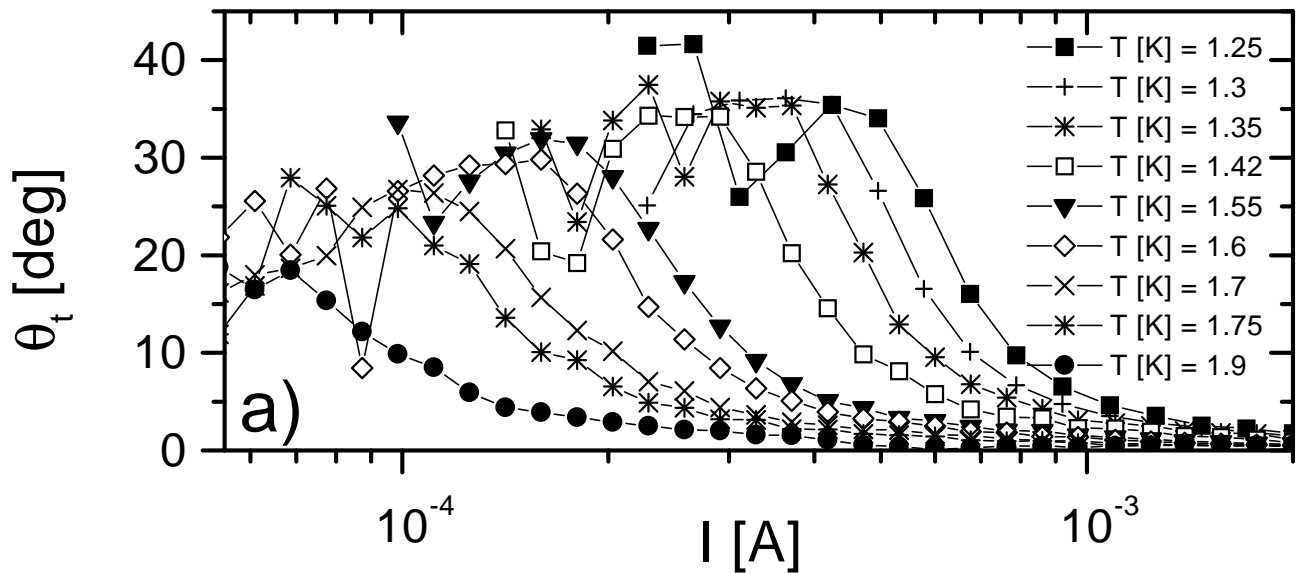
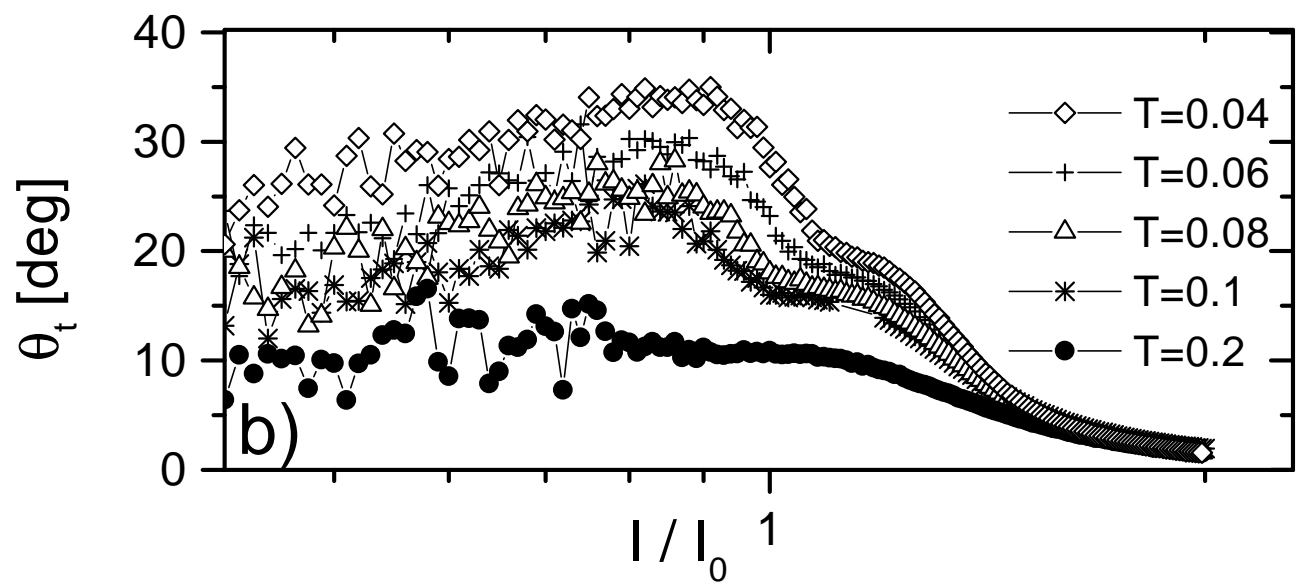
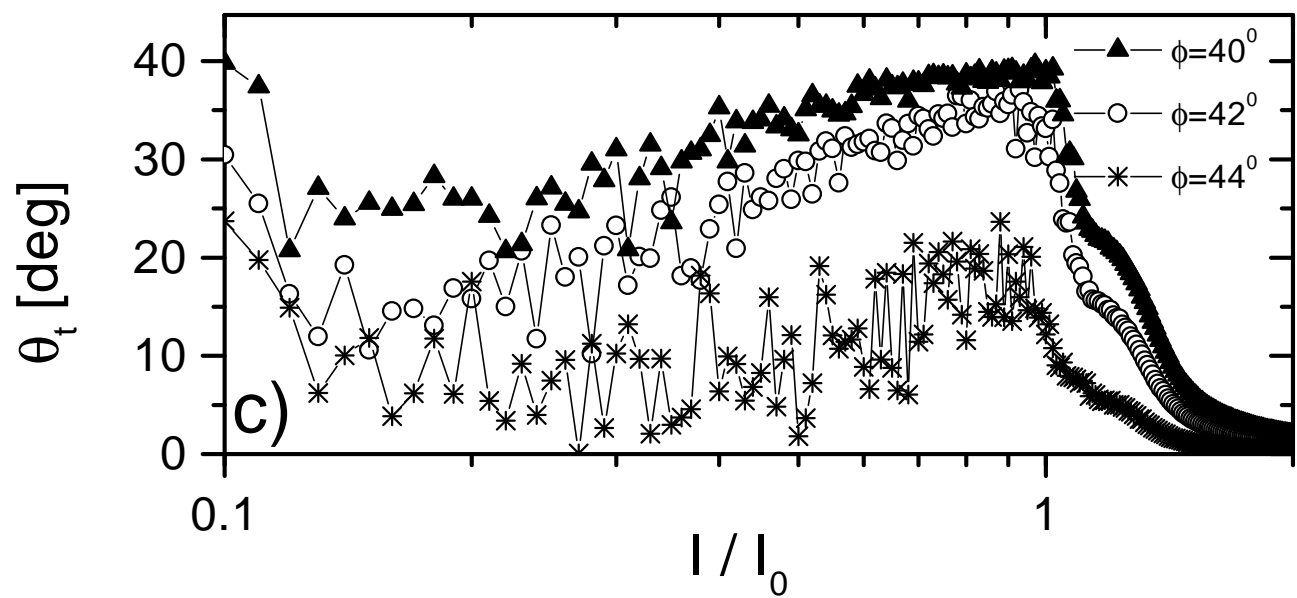
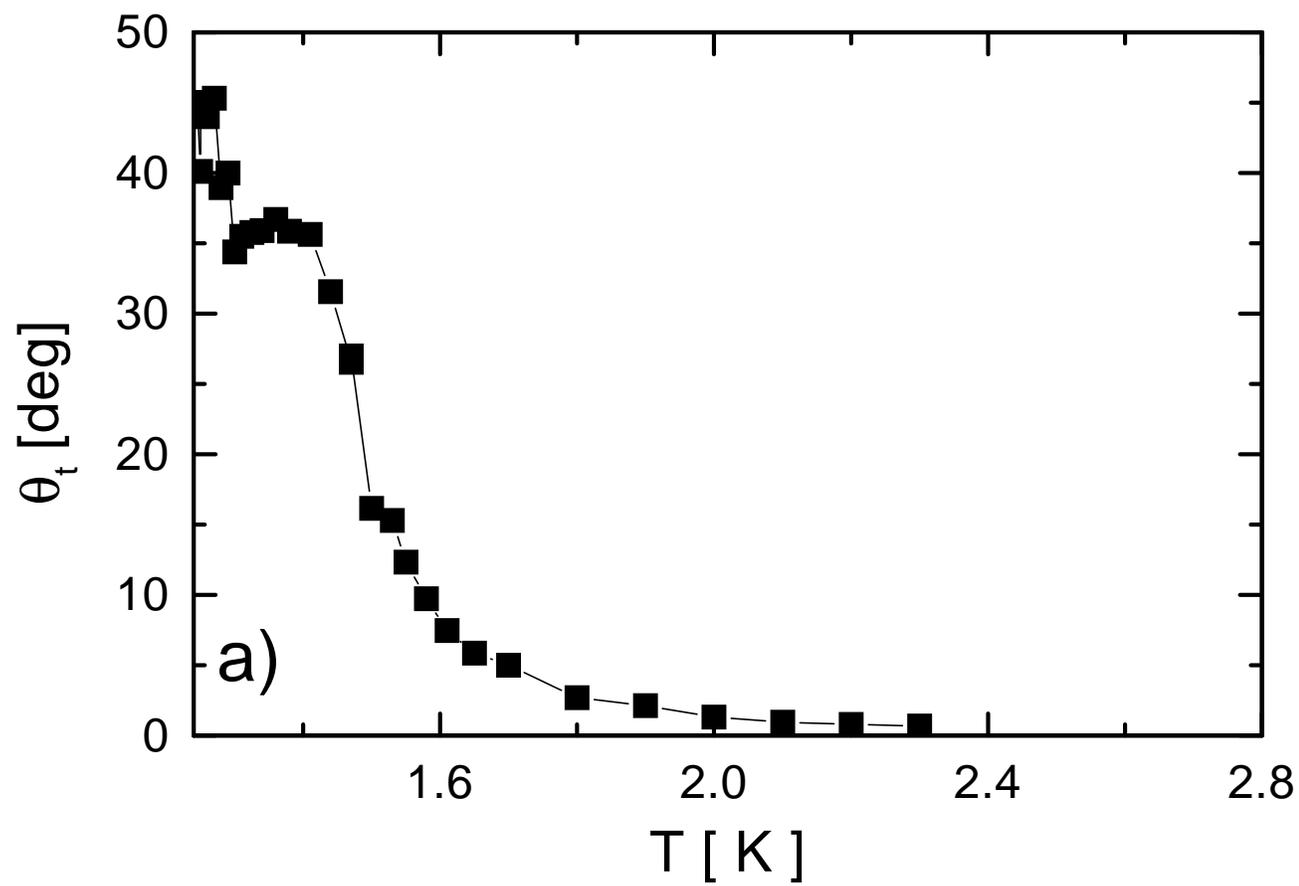
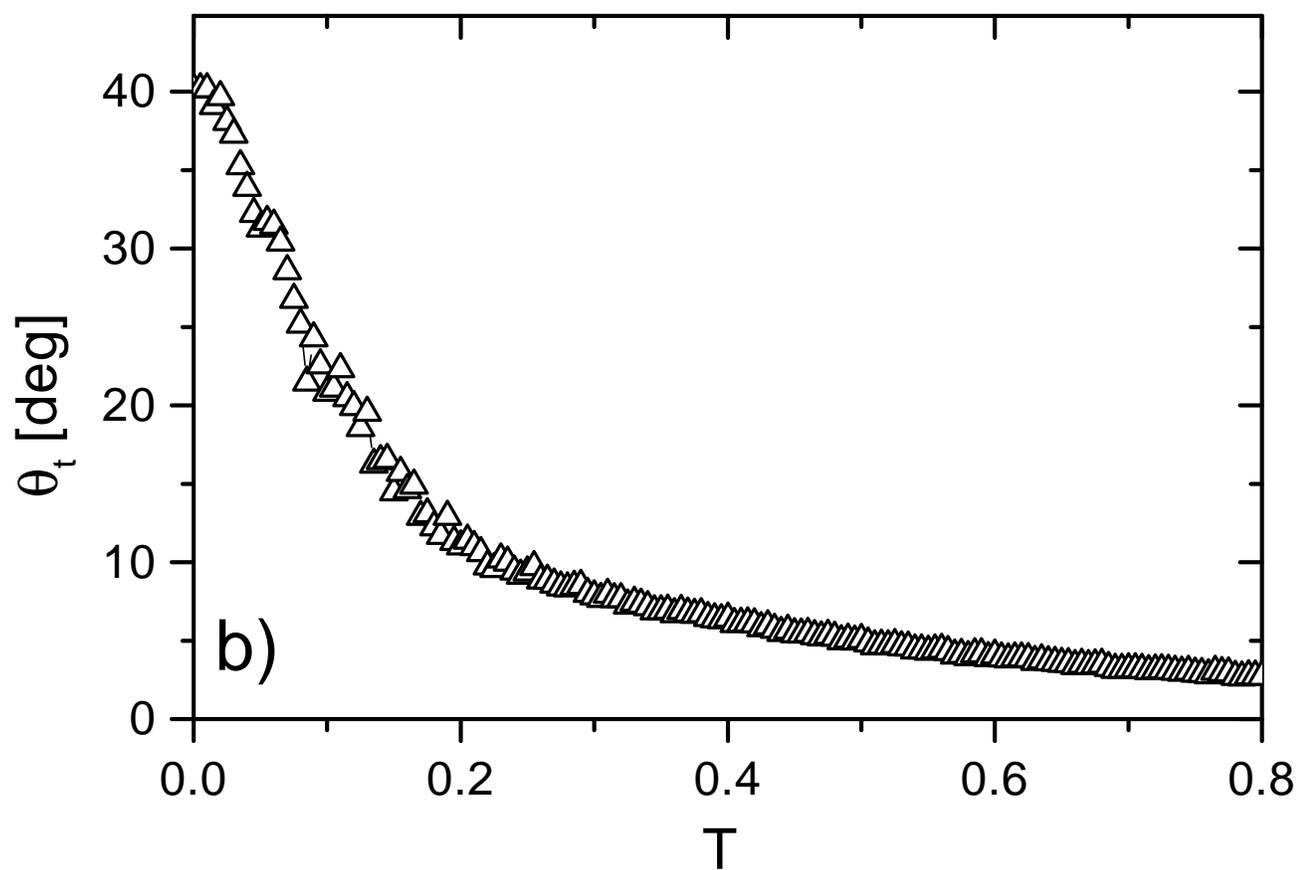
\end{document}